\RequirePackage{lineno}
\documentclass[aps, prl,
floatfix,twocolumn,superscriptaddress,
groupedaddress,
showpacs]{revtex4-1}
\pdfoutput=1
\usepackage{t1enc}
\usepackage[english]{babel}
\usepackage{graphicx}
\usepackage{epsfig}
\usepackage{amssymb}
\usepackage{dcolumn}
\usepackage{bm}
\usepackage{color}
\usepackage{float}
\usepackage{natbib}
\usepackage{footnote}
\usepackage{ulem}
\addto\captionsenglish{}
\newcommand{\BFCA}{\ensuremath{\mathrm{Ba}(\mathrm{Fe}_{1-x}\mathrm{Co}_{x})_{2}\mathrm{As}_{2}}}

\newcommand{\BFA}{\ensuremath{\mathrm{BaFe}_{2}\mathrm{As}_{2}}}
\newcommand{\KBFA}{\ensuremath{\mathrm{K}_{x}\mathrm{Ba}_{1-x}\mathrm{Fe}_{2}\mathrm{As}_{2}}}
\newcommand{\KEFA}{\ensuremath{\mathrm{K}_{x}\mathrm{Eu}_{1-x}\mathrm{Fe}_{2}\mathrm{As}_{2}}}

\newcommand{\NFCA}{\ensuremath{\mathrm{NaFe}_{1-x}\mathrm{Co}_{x}\mathrm{As}}}
\newcommand{\NFRA}{\ensuremath{\mathrm{NaFe}_{1-x}\mathrm{Rh}_{x}\mathrm{As}}}

\newcommand{\BSCCO}{\ensuremath{\mathrm{Bi}_2\mathrm{Sr}_2\mathrm{CaCu}_2\mathrm{O}_{8}}}

\newcommand{\KFA}{\ensuremath{\mathrm{KFe}_{2}\mathrm{As}_{2}}}

\begin{document}

\title{Experimental evidence for the importance of  Hund's exchange interaction for the incoherence of the charge carriers in iron-based superconductors}

\author{J.\,Fink$^{1,2,3}$, E.D.L.\,Rienks$^{1,3}$, S.\, Thirupathaiah$^4$, J. \,Nayak$^2$, A. \,van Roekeghem$^{5}$, S.\,Biermann$^{6,7}$, T.\,Wolf$^8$, P. Adelmann$^8$, H.S.\,Jeevan$^9$\altaffiliation[Present address: ] {Department of Physics, PESITM, Sagar Road, 577204 Shimoga, India},  P. \,Gegenwart$^9$, S.\,Wurmehl$^{1,3}$, C.\,Felser$^2$, B.\,B\"uchner$^{1,3}$}
\affiliation{
\mbox{$^1$Leibniz Institute for Solid State and Materials Research  Dresden, Helmholtzstr. 20, D-01069 Dresden, Germany}\\
\mbox{$^2$Max Planck Institute for Chemical Physics of Solids, D-01187 Dresden, Germany}\\
\mbox{$^3$Institut f\"ur Festk\"orperphysik,  Technische Universit\"at Dresden, D-01062 Dresden, Germany}\\
\mbox{$^4$Solid State and Structural Chemistry Unit, Indian Institute of Science, Bangalore, Karnataka, 560012, India}\\
\mbox{$^5$CEA, LITEN, 17 Rue des Matyrs, 38054 Grenoble, France}\\
\mbox{$^6$Centre de Physique Th\'{e}orique, Ecole Polytechnique, 91128 Palaiseau Cedex, France}\\ 
\mbox{$^7$LPS, Universit\'{e} Paris Sud, B\^{a}timent 510, 91405 Orsay, France}\\
\mbox{$^8$Karlsruhe Institute of Technology, Institut f\"ur Festk\"orperphysik, 76021 Karlsruhe, Germany}\\
\mbox{$^9$Institut f\"ur Physik, Universit\"at Augsburg, Universit\"atstr.1, D-86135 Augsburg, Germany}\\
}

\date{\today}

\begin{abstract}
 Angle-resolved photoemission spectroscopy (ARPES) is used to study  the  scattering rates of charge carriers from the hole pockets near $\Gamma$ in the iron-based high-$T_c$ hole doped superconductors \KBFA\ $x=0.4$ and \KEFA\ $x=0.55$ and the electron doped compound \BFCA\ $x=0.075$.   The scattering rate for any given band is found  to depend linearly on energy, indicating a non-Fermi liquid regime. The scattering rates in the hole-doped compound are considerably larger than those in the electron-doped compounds. In the hole-doped systems  the scattering rate of the charge carriers of the inner hole pocket is about three times bigger than the binding energy indicating that the spectral weight  is heavily  incoherent. The strength of the scattering rates  and the difference between electron and hole doped compounds signals the importance of  Hund's exchange coupling for  correlation effects in these iron-based high-$T_c$ superconductors. The experimental results are in qualitative agreement with theoretical calculations in the framework of combined density functional  dynamical mean-field theory. 
\end{abstract}


\maketitle

\paragraph{Introduction.} Originally  iron-based superconductors (FeSC)\,\cite{Johnston2010}  were believed to exhibit  only moderate electronic Coulomb correlations  because  X-ray absorption data derived an on-site Coulomb interaction $U$ of less than 2   eV\,\cite{Kroll2008,Yang2009b}. More recent theoretical work\,\cite{Haule2009,Aichhorn2009,Aichhorn2010,Medici2011,Werner2012,Razzoli2015} however emphasized that in contrast to the cuprates, where the correlation effects are dominated by $U$, in the  FeSCs -- because of their intrinsically multiorbital character -- there is another factor to consider: the Hund exchange interaction $J_H$.

ARPES is a suitable method to obtain information on the strength of correlation effects since it delivers the energy ($E$) and momentum ($\mathbf{k}$) dependent  self-energy function $\Sigma(E,\mathbf{k})$ from which one can derive the mass enhancement  and the  scattering rate of the charge carriers due to many-body effects\,\cite{Damascelli2003}\,\footnote{see information presented in the Supplement}. 

There are numerous  experimental studies on the mass enhancement in FeSCs using various methods which apparently support the strong influence of correlation effects, in particular related to  Hund's exchange interaction. A recent compilation of such data was  published in Ref.\,\cite{Roekeghem2016}. The effective masses show a remarkably large variance: e.g. the effective mass of \KFA\ varies between 2 and 19. There are several reasons for these uncertainties: (1) the derived effective masses are  related to the theoretical values of the bare mass, usually taken from a DFT band structure calculation, (2) there are several ARPES studies\,\cite{Ding2011,Lubashevsky2012,Starowicz2013,Fink2015,Fink2016} indicating that the  mass renormalization  is energy dependent (enhanced at low energies), which for methods covering different energy ranges, leads to different values for the effective mass.  

Recently, ARPES studies of electron doped FeSCs on the scattering rate or the lifetime broadening $\Gamma(E)$, equal to twice the imaginary part of the self-energy $\Im\Sigma(E)$,  have been presented by several groups \,\cite{Rienks2013,Fink2015,Brouet2016,Miao2016,Yi2013,Pu2016}. To our knowledge, no studies of the energy dependence of $\Gamma(E)$ in hole-doped FeSCs exist in the literature. Besides the orbital dependencies of the scattering rate, the temperature dependent crossover from  coherent quasiparticles to incoherent charge carriers is believed to provide support for a new metallic phase which was dubbed Hund's metal\,\cite{Haule2009}. Ref.\,\cite{Werner2012} theoretically predicted that the regime should be associated with a characteristic energy dependence of  the lifetimes of elementary excitations, in particular not following the parabolic behavior that is a hallmark of the Fermi liquid. Within a model context, such behavior was investigated within  high-precision renormalization group techniques in Ref.\,\cite{Stadler2015} and interpreted as  an intermediate "spin-orbitally separated" regime, where screened orbital degrees of freedom are coupled to slowly fluctuating spins that are not yet Kondo-screened. At very low temperatures, a crossover to a Fermi liquid regime is expected. On the other hand, in any normal metal there is a crossover from a Fermi liquid to an incoherent behavior at higher temperatures when the lifetime broadening $\Gamma$  exceeds the binding energy of the charge carriers. Therefore, experimental work is needed to investigate the nature of the bad metallic phase, to identify the intermediate regime, to check the theoretical predictions and to analyze the energy-dependence of the scattering rates. 

In this contribution we use ARPES to study  the scattering rates  in the hole doped "122" ferropnictides  \KBFA\ and \KEFA\ and compare the results with those derived from our  previous ARPES experiments\,\cite{Thirupathaiah2010,Thirupathaiah2011,Rienks2013,Fink2015} on electron doped compounds. The essential result of the present study is that the linear-in-energy non-Fermi-liquid scattering rates in the hole doped compounds are considerably higher than in the electron doped compounds. We ascribe the large scattering rates in the hole doped compounds, different to the cuprates, to the proximity to a compound with a $3d^5$ configuration in which, similar to Mn compounds,  Hund's exchange interaction $J_H$ is important for the correlation effects.  The ARPES results are in qualitative agreement with our  calculations in the framework of DFT combined with dynamical mean-field theory (DMFT). 
\begin{figure}[tb]
\centering
 \vspace{-2.6cm}
\includegraphics[angle=0,width=8 cm]{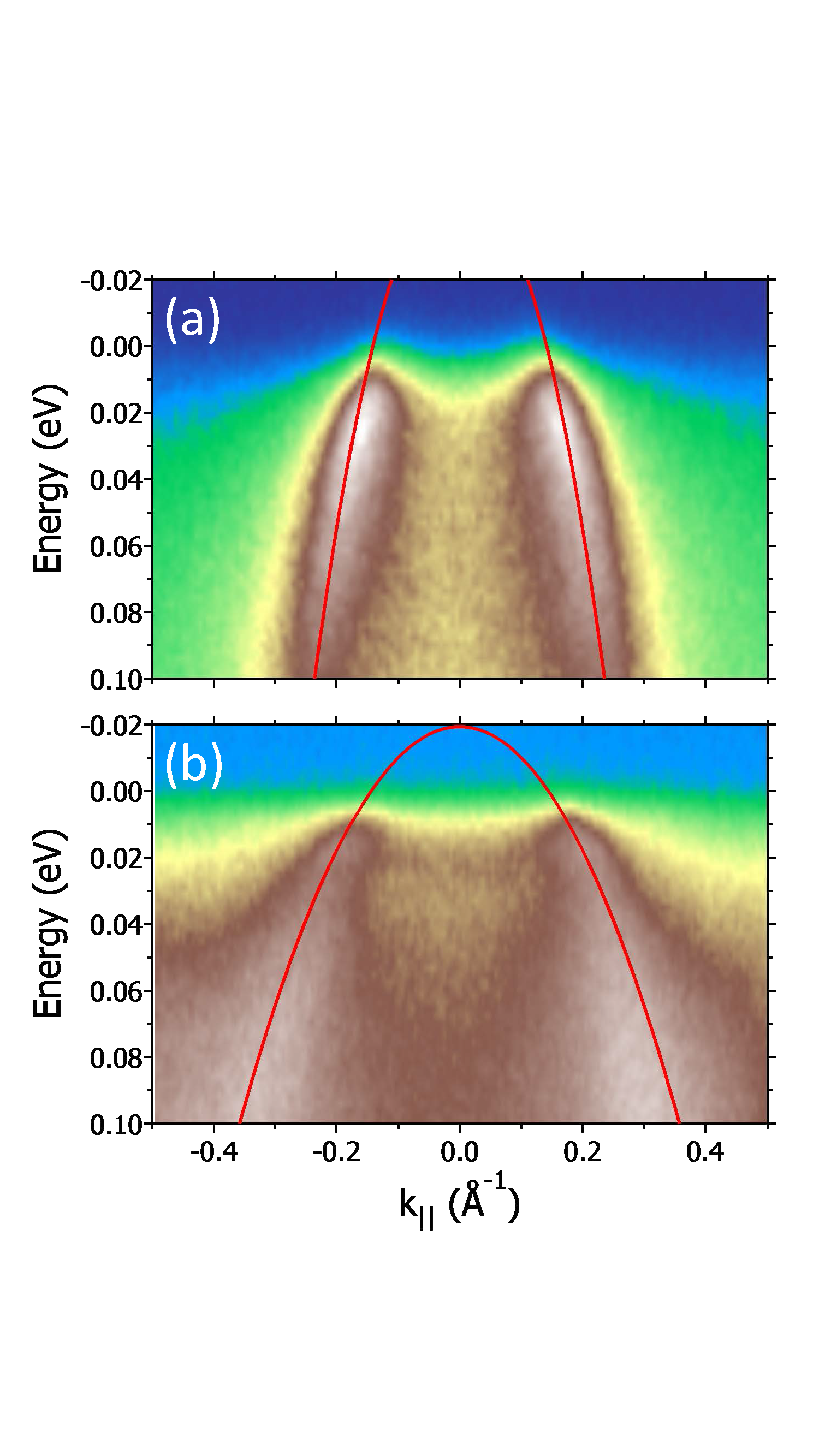}
\vspace{-2cm}
 \caption{
  (Color online) Energy distribution maps of the hole pockets of \KBFA\ $x=0.4$ near the center of the Brillouin zone, measured along the $\Gamma-M$ direction. The red lines indicate the derived band dispersion. (a) Spectral weight of the inner hole pocket measured at a temperature of $T=40$ K using vertically polarized photons with an energy of 75 eV. (b) Spectral weight of the middle hole pocket measured at $T=1.5$ K using horizontally polarized photons with an energy of 47 eV. 
  }
\label{edm}
\centering
\end{figure}

\paragraph{Experimental.}Single crystals were grown using the  self-flux technique and characterized by transport and thermal properties measurements\,\cite{Jeevan2011,Bohmer2015}.
ARPES measurements were conducted at the $1^3$-ARPES endstation attached to the  beamline
UE112 PGM 2 at BESSY with energy and angle resolutions between 4 and 15 meV and 0.2  $^\circ$, respectively. Variable photon energies h$\nu=20-130$ eV were used to reach different $k_z$ values in the Brillouin zone (BZ). The use of polarized
photons allows the selection of spectral weight with a specific  orbital character by matrix element effects\,\cite{Fink2009}. 

\paragraph{Theoretical.}
We have performed combined density functional dynamical mean field theory ("DFT+DMFT") calculations\,\cite{Lichtenstein1998,Anisimov1997} using the DFT+DMFT implementation of Ref.\,\cite{Aichhorn2009}. We have chosen the Local Density Approximation (LDA) to the exchange-correlation functional, and Hubbard and Hund's interactions obtained from the constrained random phase approximation (cRPA)\,\cite{Aryasetiawan2004} in the implementation of Ref.\,\cite{Vaugier2012}. The cRPA calculations\,\cite{Roekeghem2016a} yield $F^0=2.6$ eV, $F^2=6.2$ eV and $F^4=4.7$ eV corresponding to a Hund's rule coupling of $J_H=0.8$ eV. Calculations were performed at an inverse temperature of 100 eV$^{-1}$, corresponding to 116 K. The DMFT equations were solved using a continuous-time Quantum Monte Carlo solver\,\cite{Gull2011} as implemented in the TRIQS package\,\cite{TRIQS}, followed by an analytical continuation procedure using the maximum entropy algorithm\,\cite{Jarrell1996}.

\paragraph{Results.}
In the present contribution we focus on the scattering rates of charge carriers from hole pockets and in particular from the inner hole pocket. The reason for this is that  in the previous studies of electron-doped and P substituted compounds, $\Gamma(E)$ was found to be strongest for the inner hole pocket\,\cite{Fink2015}. Moreover the superconducting gap is largest for  the inner hole pocket\,\cite{Borisenko2012,Umezawa2012}.
In Fig. 1 we show representative data of the spectral weight of the inner (a) and the middle hole pocket (b)  near the center of the BZ  of optimally doped \KBFA\ $x=0.4$ with a superconducting transition temperature $T_c=30$ K. Along this direction the inner hole pocket has predominantly Fe 3$d_{yz}$ character, while the middle hole pocket has predominantly Fe 3$d_{xz}$ character\,\cite{Fink2015}.  The inner hole pocket is measured in the normal state to avoid the influence of the superconducting gap. For the middle hole pocket we present data in the superconducting state to show the Bogoliubov-like back dispersion near the Fermi level. The intensity of the outer hole pocket with predominantly 
Fe 3$d_{xy}$ character  is very weak in this compound  because bands with this orbital character exhibit the highest elastic scattering rates\,\cite{Herbig2015}.
\begin{figure}[tb]
\centering
 \includegraphics[angle=0,width=7 cm]{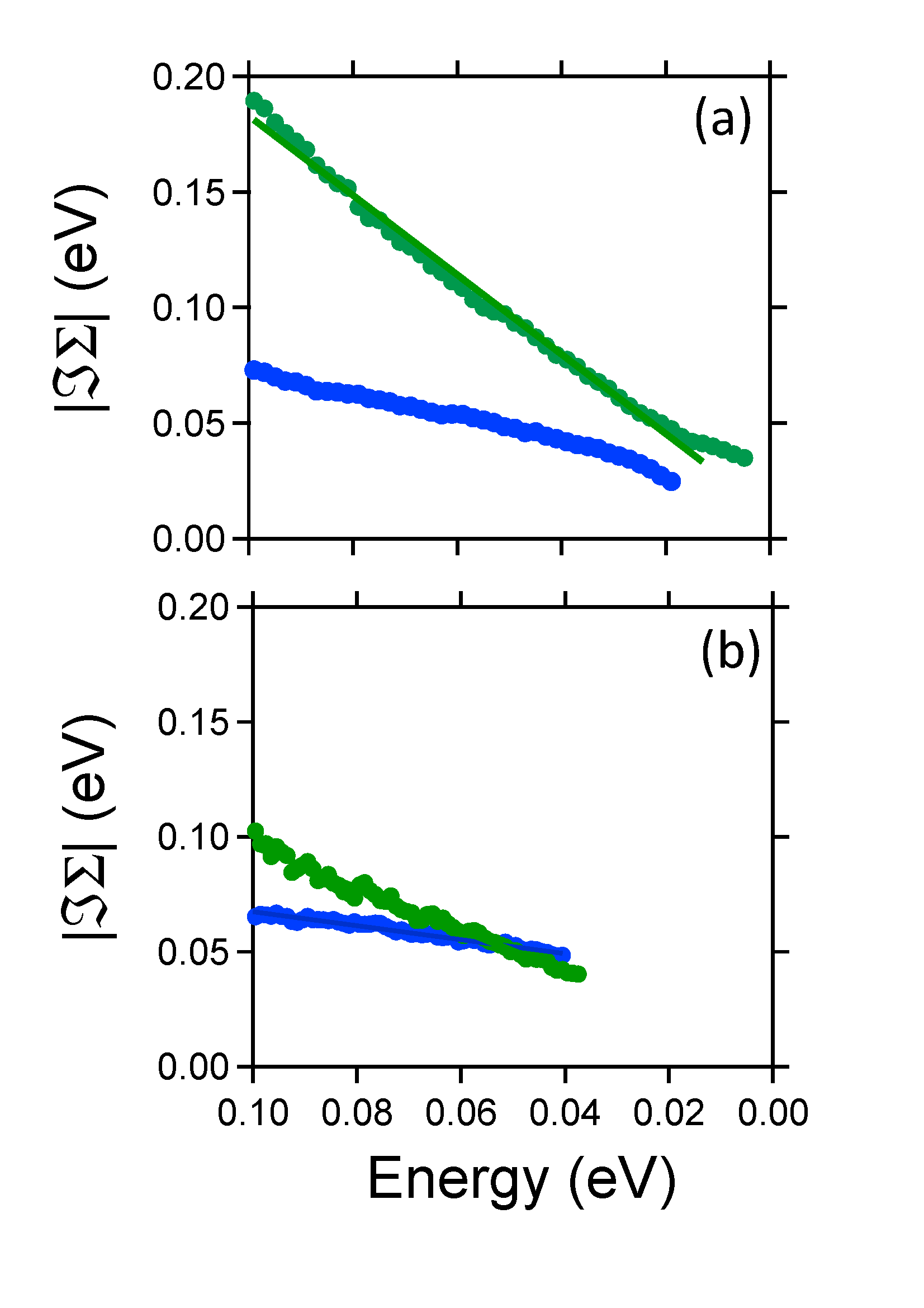}
\caption{
(Color online)   $\Im\Sigma$ as a function of binding energy for  hole pockets in hole-doped \KBFA\ $x=0.4$ (a) and electron-doped \BFCA\ $x=0.075$ (b), as derived from the ARPES spectral weight. The green data correspond to the inner hole pocket, while the blue data correspond to the middle hole pocket. The lines are derived from a linear fit. 
} 
\label{mass}
\centering
\end{figure}

Similar ARPES experiments were performed at $T$=1 K on the slightly overdoped compound \BFCA\ $x=0.075$ ($T_c= 22$ K) for the inner two hole pockets near $\Gamma$ which were already presented in  previous publications\,\cite{Thirupathaiah2010,Thirupathaiah2011,Rienks2013}. Also in this compound only the inner two hole pockets are visible at the $\Gamma$ point but not the outer one.

We have analyzed the ARPES data by a new method using a two-dimensional fit of the measured spectral function $A(E,k)$\,\cite{Note1}. In this way we derived the dispersion, approximated by a polynomial and $\Im\Sigma (E)$. We have avoided to use one of the standard evaluation methods. The first method  consists in fitting   cuts at constant energy (the momentum distribution curves MDCs)  by Lorentzians. To obtain the scattering rates $\Gamma(E)$ the   widths in momentum space are multiplied by the renormalized  velocity. As shown in the Supplement information\,\cite{Note1} the method is only correct for dispersions with small curvatures and for a small energy dependence of $\Im\Sigma$. In the second method the MDC widths are multiplied by the bare particle velocity to derive $\Im\Sigma$ for the \textit{coherent} part of the spectral weight. This is correct for systems in which the coherent spectral weight is well separated from the incoherent spectral weight and is mainly located in satellites\,\cite{Engelsberg1963,Mahan2000,Damascelli2003}. As will be shown below, in the ferropnictides as well as in other correlated systems such as the cuprates, even near the Fermi level, a large part of the spectral weight is incoherent and therefore it is only possible to derive $\Gamma(E)$ or $\Im\Sigma(E)$ from the sum of coherent and incoherent charge carriers.  Using our new evaluation method we also avoid having to  account for the bare particle velocity that is in principle unknown and which  is usually taken from DFT calculations.

In this way we obtain  $\Im\Sigma(E,k)$  for the two inner hole pockets of \KBFA\ $x=0.4$ and \BFCA\ $x=0.075$, depicted in Fig. 2. In certain energy ranges $\Im\Sigma$ can be described by a linear-in-energy relationship  $\Im\Sigma=\alpha+\beta E$. At low energy the data are limited by the finite energy resolution ($\approx 4 $ meV), by  thermal excitations for data taken at finite temperature in the normal state (see Fig. 2(a)), for measurements in the superconducting state by three times the superconducting gap $\Delta$ (for \KBFA\ $x=0.4$ $\Delta \approx 6 $ meV), and for bands which are separated from the Fermi level by a gap $E_g$ by 2$E_g$ (for the slightly overdoped \BFCA\ $x=0.075$ $E_g \approx 25$ meV). On the high-energy side, the data are limited by the crossing of other bands. The constant term $\alpha$ is due to elastic scattering of the charge carriers, e.g. by the  dopants or by contamination of the surface. The coefficient $\beta$ of the linear term, absent in a normal Fermi liquid, is  however a strong indicator for the strength of correlation effects. For the hole  doped compound we derive $\beta$ values of 1.7 and 0.5 for the inner and the middle hole pocket, respectively. Similar values for $\beta$ have been obtained for \KEFA\ $x=0.55$ ($\beta=$ 1.8 and 0.7 for the inner and the middle hole pocket, respectively). For the electron doped compound \BFCA\ we obtain $\beta$ values of 0.8 and 0.3 for the inner and the middle hole pocket, respectively. These $\beta$ values  are similar to those derived for the electron-doped compounds \NFCA\ and
\NFRA\ \cite{Fink2015}. 

\paragraph{Discussion.} The linear-in-energy  increase of $\Im\Sigma(E)$ signals no evidence for a coupling to bosonic excitations. If e.g. phonons would determine  $\Im\Sigma$,  a step like increase should be observed at the phonon energies close to 40 meV\,\cite{Engelsberg1963}. Moreover, similar to the electron doped  and the P substituted systems\,\cite{Fink2015},
no kinks are observed in the dispersion  of the hole doped compound \KBFA\ .

On the basis of phase space arguments it is easy to understand that due to the Pauli principle, in a normal Fermi liquid the scattering rate should be proportional to $E^2$\,\cite{Mahan2000}.
Various ways of reaching non-Fermi liquid regimes have been discussed in the literature. In \,\cite{Fink2016}, it was argued that the correlation induced enhanced phase space for electronic excitations could lead to a linear-in-energy increase of the scattering rates. When the height of the Fermi edge in the momentum distribution, which determines the renormalization factor $Z$ and the percentage of the coherent quasiparticles approaches zero, a marginal Fermi liquid is reached\,\cite{Varma2002,Note1} and $\Im\Sigma$ becomes linear in energy. In this case dispersions observed in ARPES experiments should not be mistaken for a quasiparticle dispersion. The same is true for the Hund's metal regime discussed below.
 
Generally the validity of the quasiparticle description was defined in the following way\,\cite{Grimvall1981} : The time dependent wave function of a photoelectron hole can be written
$\Psi(\mathbf{r} ,t) = \Psi(\mathbf{r} ) exp(-\frac{iEt}{\hbar}) exp(-\frac{\Gamma t}{\hbar})$. This leads to a wave function  which is in a certain time region similar to a free particle only when the phase $\frac{iEt}{\hbar}$ goes through many multiples of $2\pi$ before the wave function has decayed by an appreciable amount. Thus the region in which quasiparticles are defined is determined by $\frac{\Gamma}{E} \ll 2\pi$.  Using $\Gamma=2|\Im\Sigma| = 2\beta E$ and  the observed $\beta$ values between 0.5 and 1.7 shows that $\frac{\Gamma}{E}$ is between 1.0 and 3.4 which is not much smaller than $2\pi$.  Thus the charge carriers in the hole pockets are close to be incoherent or are completely incoherent, as in the hole doped compounds. Therefore it makes no sense to separate the spectral weight into coherent states (proportional to $Z$) and incoherent states (proportional to $1-Z$). It also makes no sense to compare our data with normal state transport properties, since those are related to less correlated charge carriers stemming from other sections of the Fermi surface. Nevertheless the large superconducting gap in the inner hole pocket\,\cite{Ding2008,Borisenko2012} signals that the electrons in that pocket contribute to the superconducting pairing. Finally we emphasize that the reduced slope of $\Im\Sigma$ at low energies in the upper curve of Fig.\,2\,(a) does not indicate a Fermi liquid behavior at low energies. Rather it is caused by the finite energy resolution, by an elastic scattering term $\alpha \approx 12$ meV, and a finite $\Im\Sigma$ at zero energy but at finite temperature, which in a marginal Fermi liquid model amounts at 40 K to $\approx 7$ meV\,\cite{Note1}.

It is interesting to compare the present data with results derived on the cuprates. ARPES experiments on \BSCCO\ along the nodal directed derived a $\beta$ value of 0.75 which also signals the incoherent character of the charge carriers in these compounds\,\cite{Valla1999a}. Finally calculating the mean-free path $l$ of the charge carriers from the relation $l=\frac{v}{\Gamma}$ in the FeSC's and cuprates leads to unphysical values below 1 \AA .

The difference between the scattering rate of the inner hole pocket and the middle hole pocket has been predicted by theoretical calculations\,\cite{Graser2009,Kemper2011} and were compared with experimental data for the electron doped and P substituted compounds\,\cite{Fink2015}. This difference is caused by the fact that the scattering rates between sections having the same orbital character is larger than those between sections having different orbital character. 

\begin{figure}[tb]
\centering
 \includegraphics[angle=0,width=8 cm]{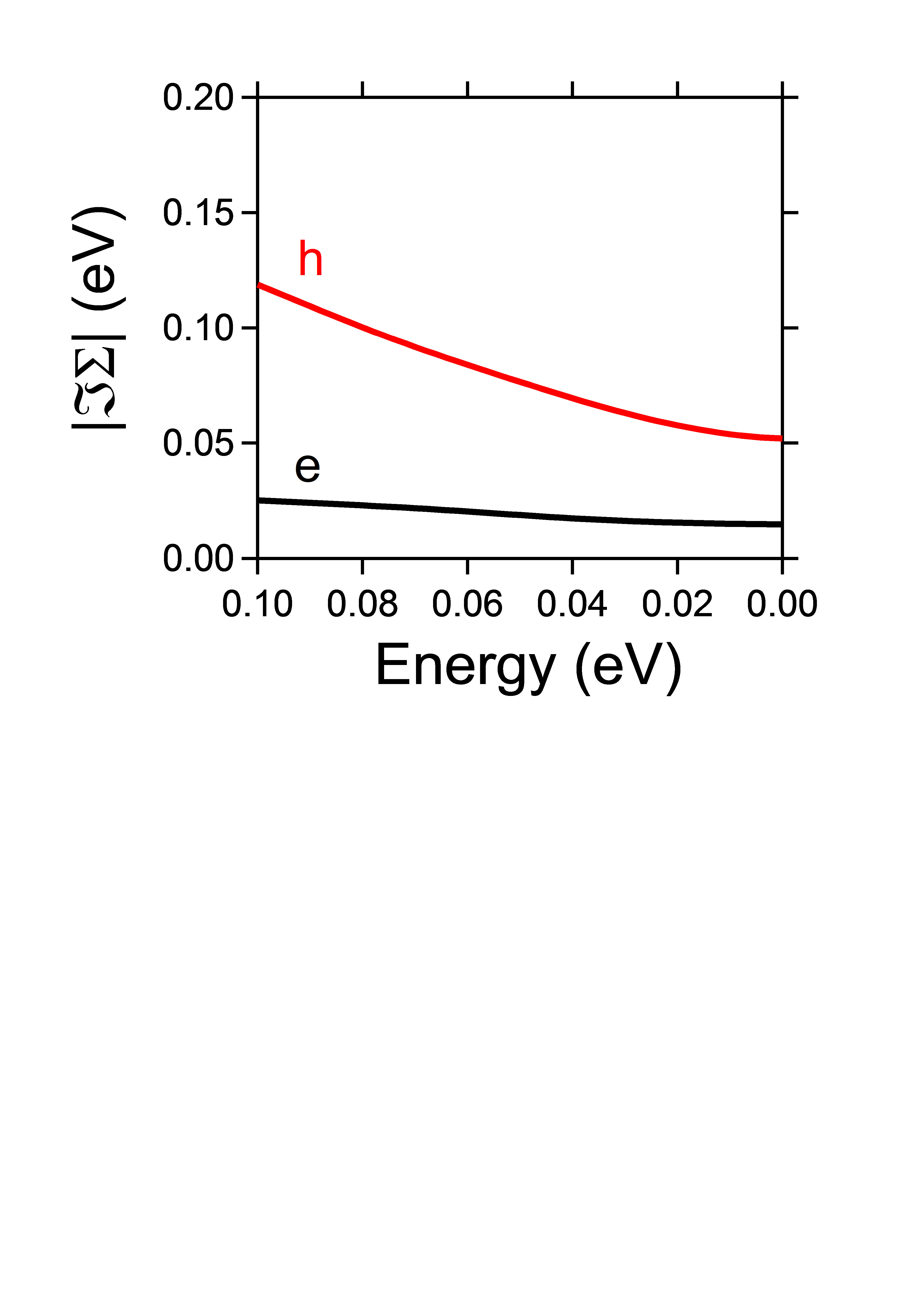}
\vspace{-5cm}
\caption{
(Color online)  DFT+DMFT calculations of   $\Im\Sigma(E)$ as a function of binding energy for  electronic states having $xz/yz$ orbital character for  \BFA\ with 0.2 holes/Fe (upper curve) and 0.075 electrons/Fe (lower curve). 
} 
\label{mass}
\centering
\end{figure}

To obtain deeper insights into the nature of the non-Fermi liquid regime in the present case and to derive an at least semi-quantitative comparison of the experimental results with theory we have performed  calculations  
in the framework of DFT+DMFT. The results are consistent with the expectation of a highly doping-dependent incoherent state\,\cite{Werner2012} based on the "spin-orbital separation" scenario\,\cite{Stadler2015} induced by Hund's coupling\,\cite{Medici2011a}: while in a half-filled system $J_H$ increases the Mott gap, the inverse is true at all other commensurate fillings\,\cite{Marel1988,Medici2015}, making the scattering rates strongly doping dependent as predicted in\,\cite{Werner2012}.
   
Results for the calculated $\Im\Sigma(E)$ are presented in Fig. 3.  For the hole-doped compound, at higher energies, an almost linear-in-energy non-Fermi-liquid behavior  with $\beta \approx 0.9$ is realized. One should notice that DFT+DMFT averages the scattering rate for a given orbital character over the hole BZ, while in the present ARPES experiment only one particular direction was analyzed, along which the orbital character of the   inner and the middle hole pocket  is dominated by  $yz$ and $xz$ states, respectively . Thus the DFT+DMFT calculation should be compared with the average value of the inner and the middle hole pocket from ARPES. The calculations for the electron-doped system yield  scattering rates which are strongly reduced. At high energies the calculations can be described by $\beta \approx 0.14$, much smaller than the measured ones.
This could indicate that the theoretical calculations still underestimate the correlation effects in these compounds. 

\paragraph{Summary.} Our study of the scattering rate of the charge carriers from the two inner hole pockets in hole-doped and electron-doped iron-based superconductors reveal a non-Fermi-liquid behavior in a large energy range. The incoherence of the charge carriers increases when going from the electron to the hole doped systems. The present experimental results underline the importance of Hund's exchange interaction for the  correlation effects in hole doped FeSCs with a $3d$ count close to five. The comparison with DFT+DMFT calculations yields almost quantitative agreement for the hole-doped compound.

This work was supported by the  German Research Foundation, the
DFG, through the priority program SPP 1458 and by the European Research Council
(Consolidator Grant 617196) and IDRIS/GENCI Orsay (project
number t2016091393). We thank Michele Casula for useful discussions.
S.T.  acknowledges support by the Department of Science and Technology (DST) through INSPIRE-Faculty program (Grant number: IFA14 PH-86).

\bibliographystyle{phaip}
\bibliography{Pnictide}

\end{document}


\title{Supplementary material for: Experimental evidence for the importance of  Hund's exchange interaction for the incoherence of the charge carriers in iron-based superconductors}

\author{J.\,Fink$^{1,2,3}$, E.D.L.\,Rienks$^{1,3}$, S.\, Thirupathaiah$^4$, J. \,Nayak$^2$, A. \,van Roekeghem$^{5}$, S.\,Biermann$^{6,7}$, T.\,Wolf$^8$, P. Adelmann$^8$, \ H.S.\,Jeevan$^9$\altaffiliation[Present address: ] {Department of Physics, PESITM, Sagar Road, 577204 Shimoga, India},  P. \,Gegenwart$^9$, S.\,Wurmehl$^{1,3}$, C.\,Felser$^2$, B.\,B\"uchner$^{1,3}$}
\affiliation{
\mbox{$^1$Leibniz Institute for Solid State and Materials Research  Dresden, Helmholtzstr. 20, D-01069 Dresden, Germany}\\
\mbox{$^2$Max Planck Institute for Chemical Physics of Solids, D-01187 Dresden, Germany}\\
\mbox{$^3$Institut f\"ur Festk\"orperphysik,  Technische Universit\"at Dresden, D-01062 Dresden, Germany}\\
\mbox{$^4$Solid State and Structural Chemistry Unit, Indian Institute of Science, Bangalore, Karnataka, 560012, India}\\
\mbox{$^5$CEA, LITEN, 17 Rue des Matyrs, 38054 Grenoble, France}\\
\mbox{$^6$Centre de Physique Th\'{e}orique, Ecole Polytechnique, 91128 Palaiseau Cedex, France}\\ 
\mbox{$^7$LPS, Universit\'{e} Paris Sud, B\^{a}timent 510, 91405 Orsay, France}\\
\mbox{$^8$Karlsruhe Institute of Technology, Institut f\"ur Festk\"orperphysik, 76021 Karlsruhe, Germany}\\
\mbox{$^9$Institut f\"ur Physik, Universit\"at Augsburg, Universit\"atstr.1, D-86135 Augsburg, Germany}\\
}

\date{\today}

\begin{abstract}
 In this Supplement we provide additional information on the evaluation of the ARPES spectra using a two-dimensional fit of the measured spectral function, depending on the energy and momentum.

\end{abstract}

\pacs{74.25.Jb, 74.70.Xa, 79.60.-i}

\maketitle

ARPES measures the energy ($E$) and momentum ($\mathbf{k}$) dependent spectral function $A(E,\mathbf{k})$  multiplied by a transition matrix element, the Fermi function, and a resolution function\,\cite{Damascelli2003}. According to Mahan\,\cite{Mahan2000}  the spectral function is given by 
\begin{eqnarray}
A(E,\mathbf{k})=-\frac{1}{\pi}\frac{\Im\Sigma(E,\mathbf{k})}
{[E -\epsilon _{\mathbf{k}}-\Re\Sigma(E,\mathbf{k})]^2+\Im\Sigma(E,\mathbf{k})^2},
\label{26_1}
\end{eqnarray}
where $\epsilon _{\mathbf{k}}$  is the bare particle energy, $\Re\Sigma(E,\mathbf{k})$ is the real part  and $\Im\Sigma(E,\mathbf{k} ) $ is the imaginary part of the self-energy. 

For a weakly correlated electron liquid, i.e., when  $\Im\Sigma(E,\mathbf{k})$ is much smaller compared to the binding energy, the maxima of the spectral function (the dispersion) are determined by the equation
\begin{eqnarray}
E -\epsilon _{\mathbf{k}}-\Re\Sigma(E,\mathbf{k})=0.
\end{eqnarray}
In the case when $\Re\Sigma(E,\mathbf{k})=-\lambda E$  as in a Fermi liquid close to the chemical potential and when $\Re\Sigma(E,\mathbf{k})$ is independent of the momentum, the dispersion follows $E=\epsilon _{\mathbf{k}}/(1+\lambda)$.
Introducing $Z=1/(1+\lambda)=m/m^*$, defines the renormalization constant Z, which is a measure of the the coherent fraction  of the spectral function relative to the total spectral weight. This relation also determines the mass enhancement of the quasi-particles $m^*/m$.  With increasing coupling constant $\lambda$ or with increasing effective mass enhancement the maximum of the spectral function is shifted to lower binding energy.  

To determine the coherent part of the self-energy in weakly interacting systems, the bare particle energy is replaced by
$v_k k$ where $v_k$ is the bare particle dispersion. In this way the spectral function at constant energy (which is called a momentum distribution curve MDC) is a Lorentzian and the maximum of this Lorentzian determines the renormalized dispersion $\epsilon^*=\epsilon/(1+\lambda)$. The full width at half maximum W multiplied by the \textit{bare particle} velocity determines the life time broadening or the scattering rate  $\Gamma=2\Im\Sigma$. This type of evaluation  is strictly applicable only for weakly correlated systems and a linear bare particle dispersion.  There are, however, numerous examples in the literature where this method was also applied for highly correlated materials and a non-linear dispersion.

In a second common evaluation method the MDCs are fitted by Lorentzians and $\Gamma(E)$ is derived by multiplying the width in momentum space with the \textit{renormalized}
velocity. We will show below that also this evaluation method is exact only in particular cases.

\begin{figure} [!htb]
  \centering
	\vspace{-1cm}
\includegraphics[angle=0,width=1.1\linewidth]{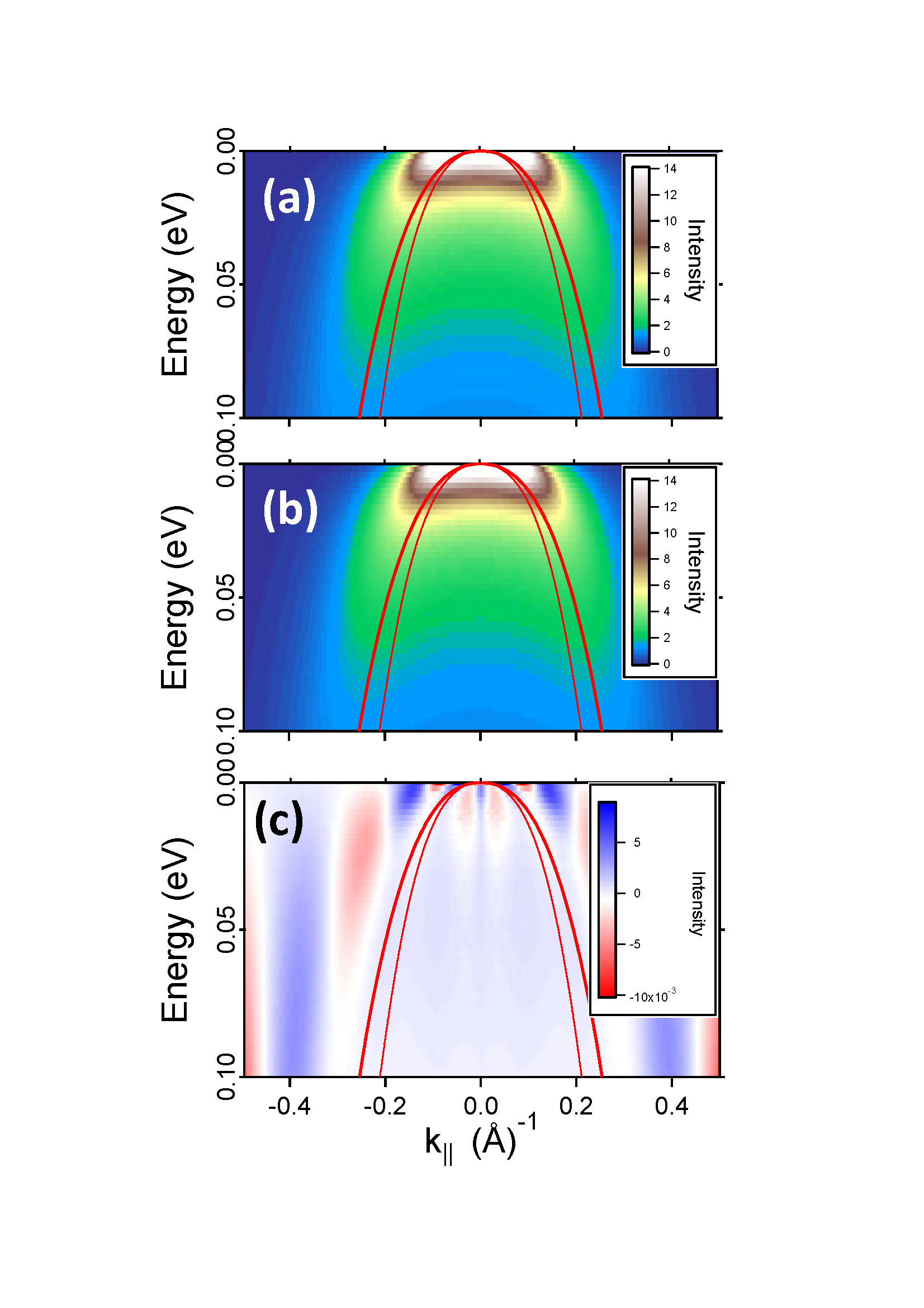}
\vspace{-1.5cm}
  \caption{(Color online) Spectral function for a marginal Fermi liquid with parameters given in the text. (a) calculated spectral function using Eq. (1).  (b) Two-dimensional fit to the data in (a) using a variable renormalized dispersion and $\Im\Sigma$. (c) Difference between the calculated and the fitted spectral function. Note that the scale is reduced by a factor of 100 compared to (a) and (b). The solid red line depicts the fitted renormalized dispersion. The thin red line shows the dispersion derived from a fit of the MDCs with two Lorentzians.} 
  \label{Specfunc}
 \centering
\end{figure}

\begin{figure} [!htb]
  \centering
	\vspace{-1.0cm}
\includegraphics[angle=0,width=\linewidth]{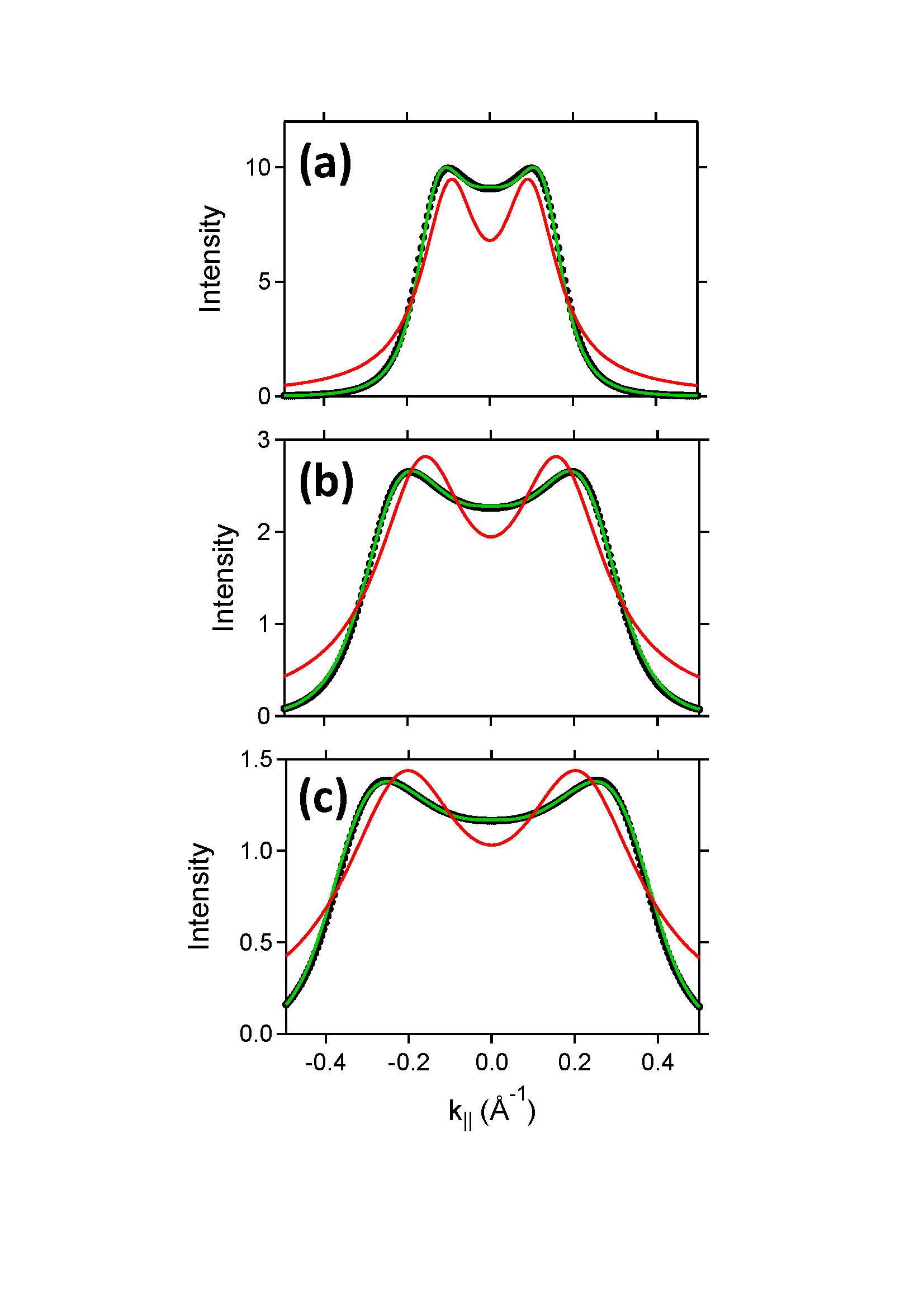}
\vspace{-2.0cm}
  \caption{(Color online) MDCs at various binding energies $E$. The black dots represent the calculated MDCs. The green curve is derived from the two dimensional fit of the spectral function. The red curve depicts  a curve derived from a fit of a sum of two Lorentzians to the calculated spectral function. (a) $E=0.01$ meV. (b) $E=0.05$ eV. (c) $E=0.1$ eV.} 
  \label{MDC}
 \centering
\end{figure}

In this section we develop a new method for the evaluation of the imaginary part of the self-energy for highly correlated systems and a non-linear dispersion. Furthermore  we discuss the errors which may appear when  the traditional methods are used. We propose to evaluate the measured spectrum by performing a two-dimensional fit using Eq (1). The fit  parameters are the values of the imaginary part of the self-energy for the values of the binding energy used in the experiment and the renormalized dispersion as a function of energy $E$, approximated by a polynomial. The latter approximation is only useful for dispersions without kinks. In case of dispersions with kinks,  a more complicated description of the dispersion is needed. Using the new method one can derive $\Im\Sigma$ also for highly correlated systems in which the spectral weight can no longer be separated into a coherent and an incoherent part. Moreover the method can also be applied for data in which   a non-linear dispersion is detected. Furthermore  this method   can avoid the use of the not well defined bare particle dispersion, which is usually taken from DFT calculations.

\begin{figure} [!htb]
  \centering
\includegraphics[angle=-90,width=\linewidth]{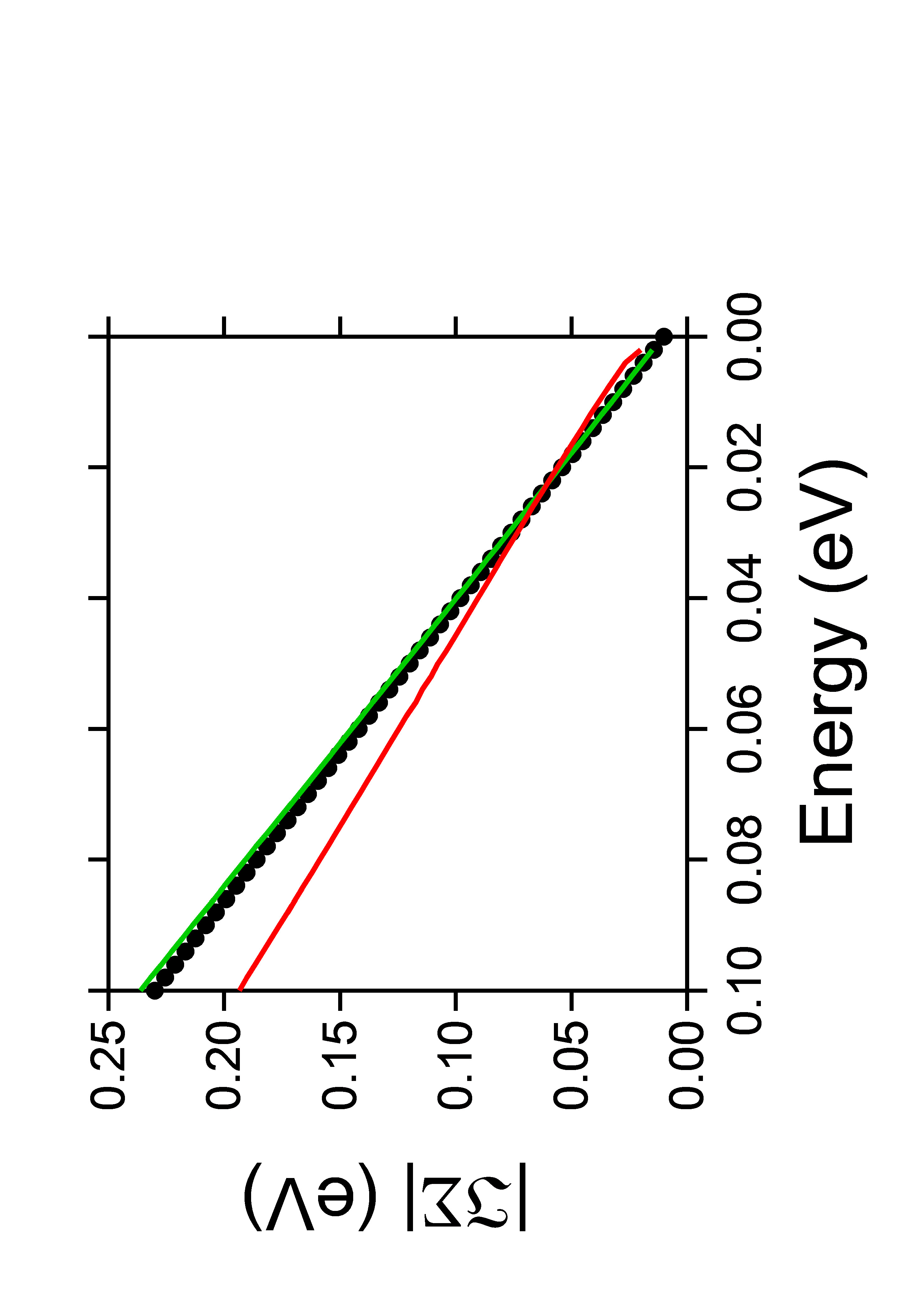}
  \caption{(Color online) Black dots: calculated $\Im\Sigma$ using Eq. 3 with the parameters given in the text. Green line:  $\Im\Sigma$ derived from the 2D fit to the calculated spectral function. Red line: $\Im\Sigma$ derived from the width of the Lorentzians multiplied by the velocity.}
 \label{ImSig}
\end{figure}
\begin{figure} [!htb]
  \centering
\includegraphics[angle=0,width=\linewidth]{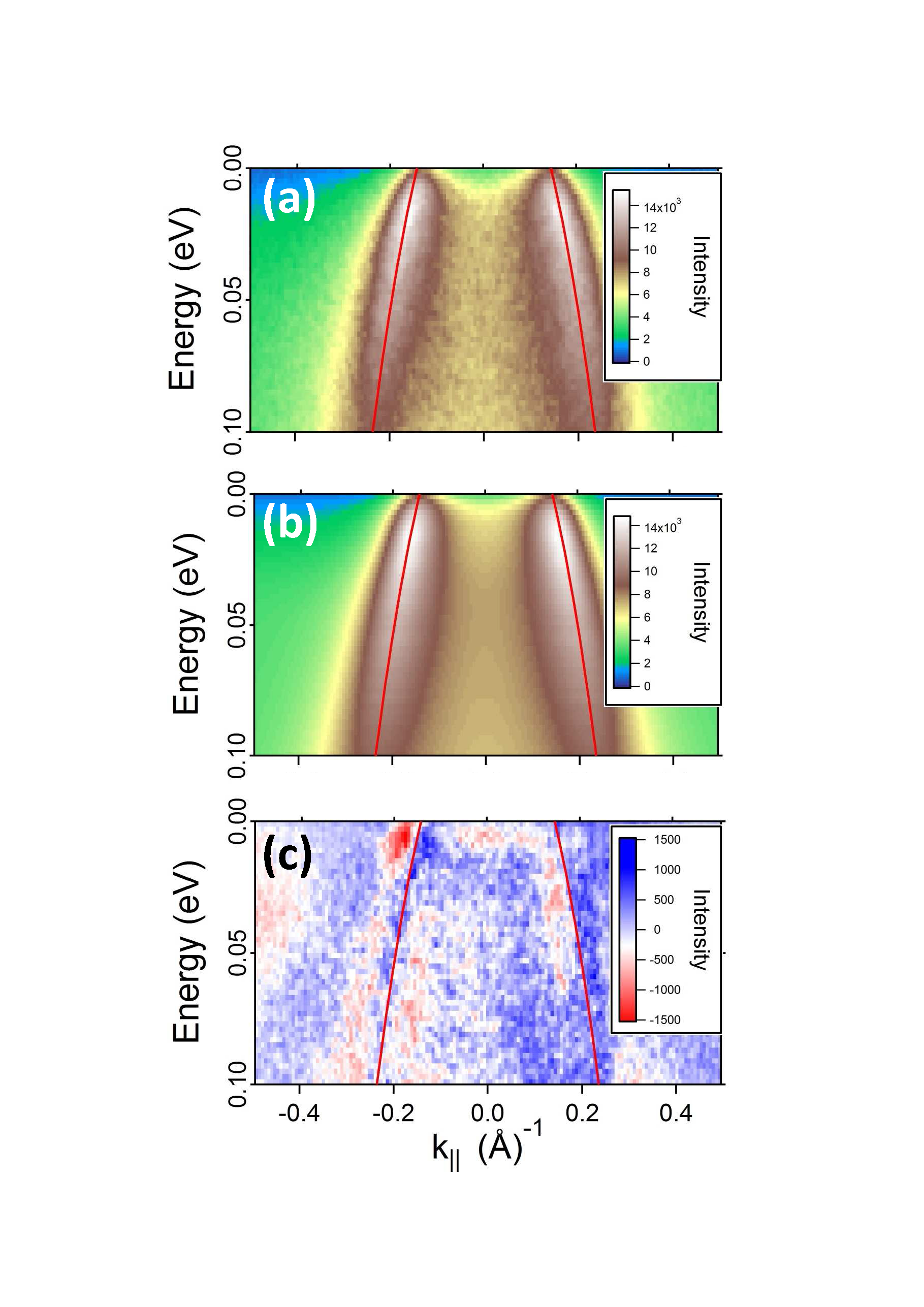}
\vspace{-1.5cm}
  \caption{(Color online) Analogous data as shown in Fig.1 but for real ARPES data of \KBFA\ .} 
  \label{Specfunc}
 \centering
\end{figure}
\begin{figure} [!htb]
  \centering
\includegraphics[angle=0,width=\linewidth]{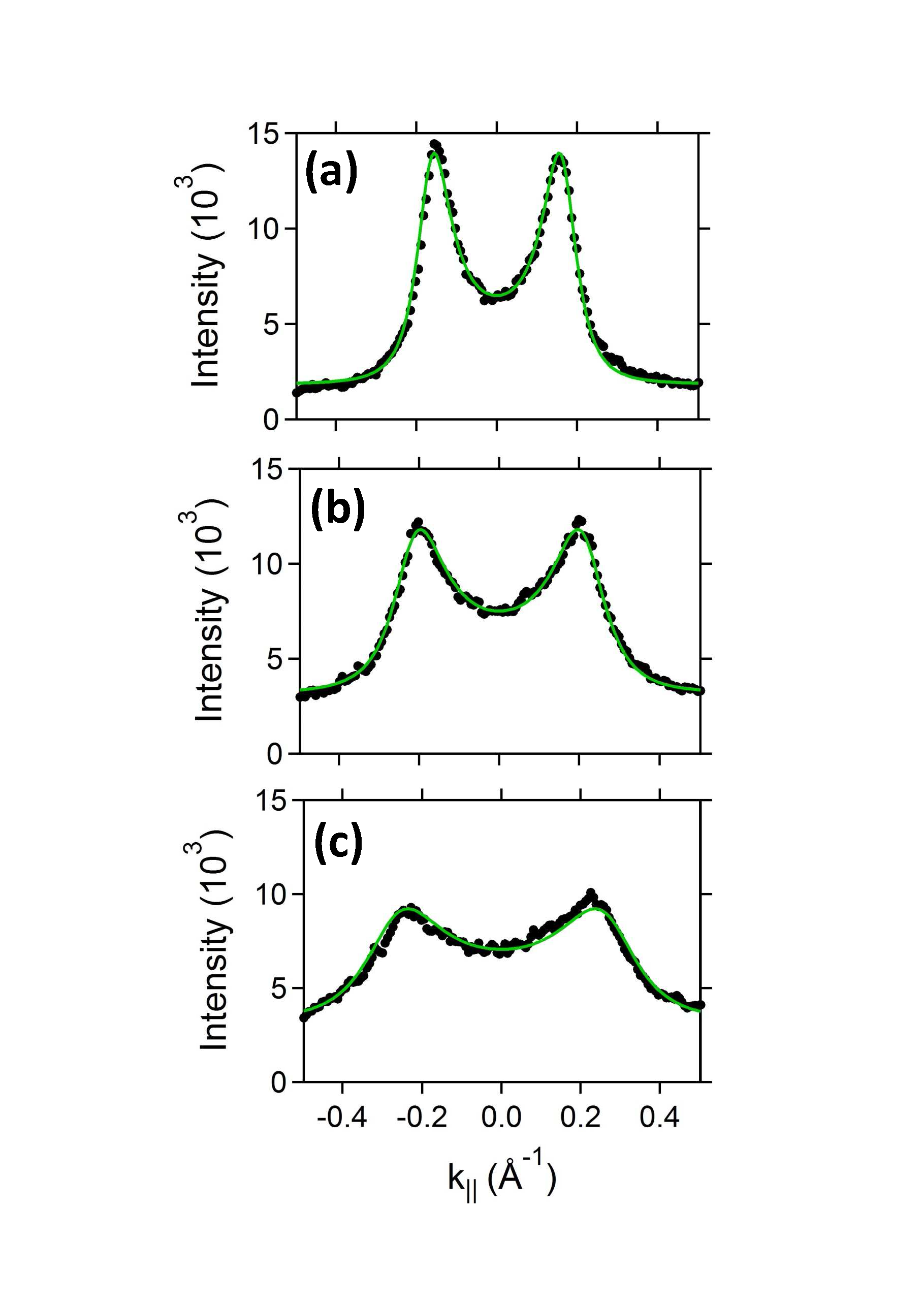}
  \caption{(Color online) Analogous data as shown in Fig.2 but for real ARPES data of \KBFA\ $x=0.4$.} 
  \label{Specfunc}
 \centering
\end{figure} 

First we test the method using a calculated spectrum, using the self-energy function from a 
 "marginal" Fermi liquid in which the coherent part of the spectral weight approaches zero ($Z=0$),  a concept originally devised to phenomenologically describe the normal state of the  cuprate superconductors.  One assumes that   the  scattering rates at low energy are much higher compared to a normal Fermi liquid due to strong correlation effects and/or nesting. The self-energy is given by\,\cite{Varma2002} 
\begin{eqnarray}
\Sigma=\frac{1}{2}\lambda_{MF}E ln\frac{E_c}{u}-i\frac{\pi}{2}\lambda_{MF}u,
\end{eqnarray}
where $u=max(|E|,kT)$,  $kT$ is the thermal energy,  $\lambda_{MF}$ is a coupling constant, and $E_c$ is an ultraviolet cutoff energy. At low temperatures the imaginary part of the self-energy is linear in energy. The  $\Re\Sigma$  is
no longer  linear in energy as in a Fermi liquid and one has to solve Eq. (2)  numerically to derive the dispersion. In the calculations we use the very high coupling constant $\lambda_{MF}=1.4$ corresponding to a slope $\beta=2.2$ of $\Im\Sigma(E)$ \textit{vs.} $E$.  The high coupling constant was used to demonstrate the difference between the new evaluation method and the standard evaluation method using the Lorentzian width and multiplying it with the renormalized velocity. In the calculation we also used  a parabolic bare particle dispersion and a cutoff energy of $E_c=1.5$  eV. The calculated spectral function together with the dispersion derived from Eq. (2) and a dispersion derived from MDC fits with Lorentzians  is shown in Fig. 1(a). 

 As shown previously\,\cite{Fink2015} for these high $\lambda$ values the renormalized dispersion at low energies is no longer parabolic due to the logarithmic term in Eq. (3).
The spectral function derived from a 2D fit to the calculated spectral function is shown in Fig. 1(b). The difference between the calculated and the fitted spectral function is shown in Fig. 1(c).
\begin{figure} [!ht]
  \centering
\includegraphics[angle=-90,width=\linewidth]{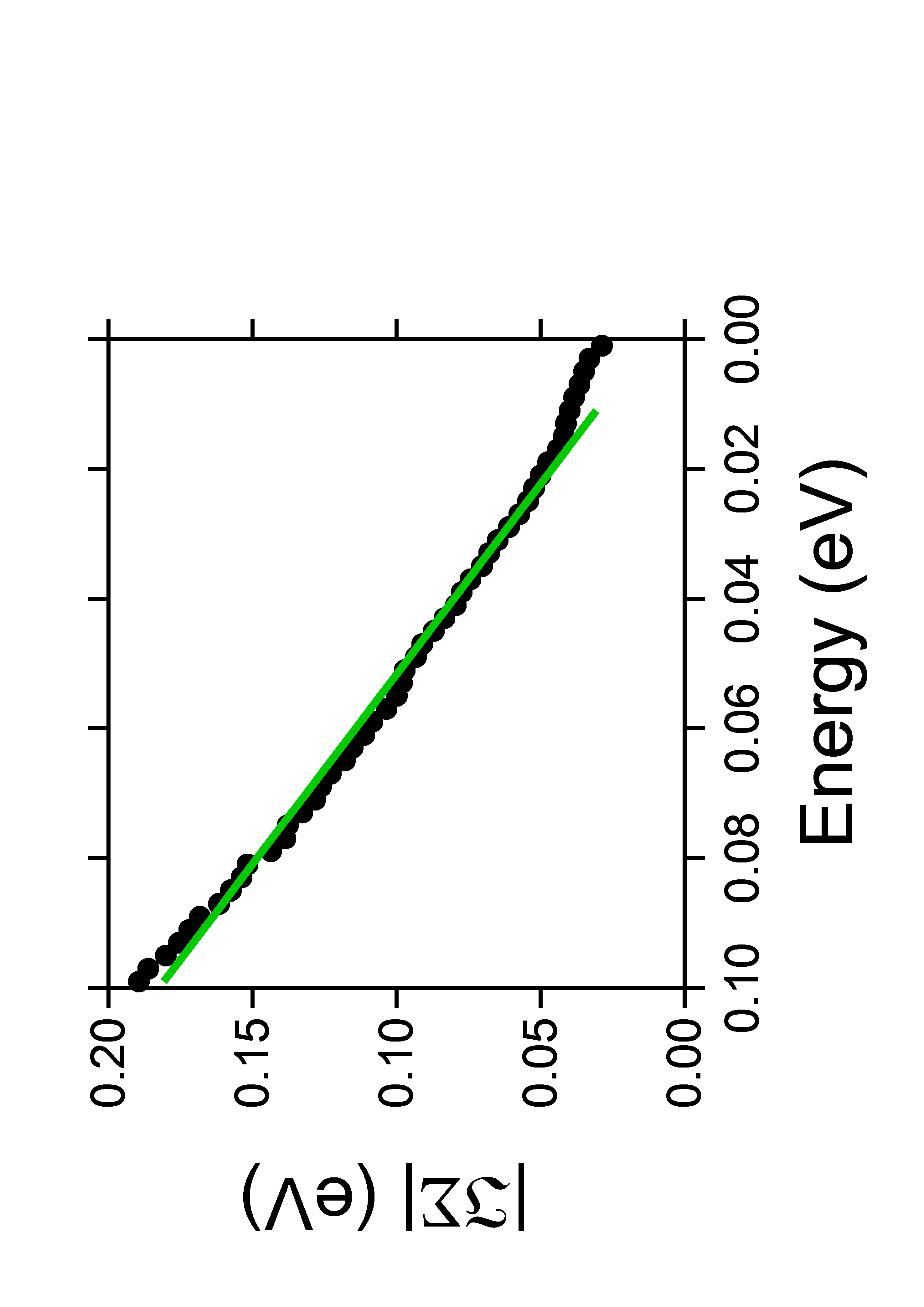}
  \caption{(Color online) Analogous data as shown in Fig. 3 but for real ARPES data of \KBFA\ $x=0.4$. } 
  \label{Specfunc}
 \centering
\end{figure}

As shown in Fig. 2 for various binding energies the MDCs of the calculated spectral function are no more a sum of two Lorentzians. Rather an asymmetric distribution determines the MDCs. An expansion of the dispersion about the $k_M$, the momentum at which the spectral function has a maximum, into a Taylor series and using a parabolic renormalized dispersion yields that the asymmetry is only zero when the renormalized velocity $v^*$  is much bigger than $\frac{\hbar}{m^*} \frac{W}{4}$.  In addition one needs  $\beta$ values which are much smaller than one to obtain a Lorentzian. Details of the new method will be published in a forthcoming publication.
The derived deviations from a Lorentzian indicate that the standard evaluation of the lifetime broadening is not exactly working for data with a non-linear dispersion and/or with a slope of $\Im\Sigma$ which is not much smaller than one.

In Fig 3 we show $\Im\Sigma(E)$ calculated from Eq. (2), derived from the 2D fit, and derived from a fit of the MDCs by Lorentzians. While the 2D fit perfectly agrees with the calculated values, there is a difference of up to 17 \% to the values derived from the fit with Lorentzians. The difference is reduced with increasing effective mass and decreasing coupling constant. 

Next we illustrate the new method by showing the detailed evaluation of real ARPES data.
We select the spectral function of the inner hole pocket of \KBFA\ $x=0.4$ shown in Fig. 1 of the main paper.
 In Fig. 4 we show the analogous data as in Fig. 1. In real ARPES data the amplitude of the spectral function is no more constant. Thus we have introduced  an energy depending decrease of the amplitude in the fit, described by two fit parameters. Furthermore we have taken into account a k-independent background which slightly increases with increasing energy.

The fitted spectral function [see Fig 4(b)] is very close to the measured spectral function [see Fig 4(a)], which is also seen in the difference between the ARPES spectral function and the fitted spectral function [see Fig 4(c)]. The MDCs, presented in an analogous way as in Fig. 2 (see Fig. 5) are well described by the fitted MDCs. Finally, we present  $\Im\Sigma$ derived from the fit in Fig. 6.
Fitting the derived results by $\Im\Sigma=\alpha+\beta E$ yields $\alpha=0.01$ eV and $\beta=1.7$.

\bibliographystyle{phaip}
\bibliography{Pnictide}